\documentclass[epjST]{svjour}
\usepackage{graphics}
\usepackage{amsmath}
\usepackage{xcolor}
\begin{document}
\title{ \textcolor{black}{Tan's contact of a harmonically trapped one-dimensional Bose gas: strong-coupling expansion and  conjectural approach at arbitrary interactions}}
\author{Guillaume Lang\inst{1}\fnmsep\thanks{\email{guillaume.lang@lpmmc.cnrs.fr}} \and Patrizia Vignolo \inst{2} \and Anna Minguzzi\inst{1}}
\institute{Universit\'e Grenoble-Alpes, LPMMC, F-38000 Grenoble, France and CNRS, LPMMC, F-38000 Grenoble, France \and  Universit\'e C\^ote d'Azur, CNRS, Institut Non Lin\'eaire de Nice,
1361 route des Lucioles,
06560 Valbonne, France}
\abstract{We study  \textcolor{black}{Tan's contact}, i.e. \textcolor{black}{the coefficient of} the high-momentum tails of the momentum distribution \textcolor{black}{at leading order}, for an interacting one-dimensional Bose gas subjected to \textcolor{black}{a} harmonic confinement. Using a strong-coupling systematic expansion of the ground-state energy of the homogeneous system stemming from the Bethe-Ansatz solution,
together with the local-density approximation, we obtain the strong-coupling expansion for \textcolor{black}{Tan's contact} of the harmonically trapped gas. Also, we use a very accurate conjecture
for the ground-state energy of the homogeneous system to obtain an approximate expression for \textcolor{black}{Tan's contact} for arbitrary interaction strength, thus estimating the accuracy of the strong-coupling expansion. Our results are relevant for ongoing experiments with ultracold atomic gases.} 
\maketitle
\section{Introduction}
\label{intro}

Continuous experimental progresses in trapping and cooling atomic gases have led to the realization of one-dimensional (1D) geometries, the observation of the effect of quantum fluctuations \cite{shlyapnikov-hannover,westbrook-palaiseau}, and the reach of the strongly correlated, Tonks-Girardeau regime \cite{Weiss,Bloch}, where many of the properties of bosons are the same as those of free fermions \cite{Girardeau}. Several physical properties have been experimentally studied, e.g. the density profiles \cite{Weiss}, the two- and three-body correlation functions \cite{Kinoshita2005,Haller}, the collective excitation spectrum \cite{Naegerl,Fang}, and transport of an impurity in a 1D gas \cite{Koehl} (see e.g. \cite{CitroRMP} for a comprehensive review). 

One of the most common experimental observables is the momentum distribution. This quantity embeds the one-body properties of the gas, in particular it corresponds to  the Fourier transform of the one-body density matrix, yielding information on the first-order spatial coherence of the gas. At zero temperature, an interacting 1D Bose gas is predicted to display quasi-off-diagonal long-range order, i.e. an algebraic decay of the one-body density matrix at large distances \cite{Haldane}. High-precision measurements of the momentum distribution are becoming available \cite{Bouchoule}, calling for new theoretical developments. 

Recent experimental studies allow to access in particular the high-momentum region of the momentum distribution \cite{Clement,Hadzibabic}. 
The high-momentum tails of the momentum distribution $n(k)$ of a gas with contact interactions display a universal $n(k)\propto 1/k^4$ decay \cite{Vignolo2002,Olshanii}, which originates from the zero-range interaction potential:
bosons with contact interactions have a cusp in the many-body wavefunction whenever the relative distance of each pair of particles vanishes. This implies a non-analyticity in the one-body density matrix at short distances, in the form $\rho_1(x)\sim |x|^3$, which leads to the high-momentum tails.

The weight of the momentum distribution tails, known as \textcolor{black}{Tan's contact} , is an important two-body quantity, related to various physical observables, such as the interaction energy, the two-body correlation function at zero distance, the rf-spectroscopy, through Tan's relations \cite{Tan1,Tan2,Tan3,Braaten,Braatenbis,Braatenter}. 
For a homogeneous 1D Bose gas, \textcolor{black}{Tan's contact}  at zero temperature can be obtained exactly using the Bethe Ansatz solution for the ground-state energy \cite{LiebLiniger,Olshanii,Lang2016}. In the experiments, ultracold atomic gases are kept together by an external trap, which is in most cases \textcolor{black}{a} harmonic confinement. \textcolor{black}{Tan's contact} of an interacting one-dimensional Bose gas under harmonic confinement has been first studied in Ref.~\cite{Olshanii}, where a numerical solution based on the local-density approximation (LDA) of the homogeneous-system result has been provided. Analytically, again in the local-density approximation, a strong-coupling expansion has been derived to zeroth order \cite{Olshanii}, as well as corrections to first and second order \cite{Decamp2016}. A comparison with matrix-product state simulations has shown that the local-density approximation works surprisingly well \cite{Decamp2016}, even for a small number of particles.

The study of the equation of state of an interacting 1D Bose gas has received a renewed attention \cite{Ristivojevic,Lang2016}. In particular, in Ref.~\cite{Lang2016} we have derived a strong-coupling expansion for the ground-state energy to an unprecedented accuracy as well as a conjectural expression which is extremely close to the exact numerical solution for a wide range of interaction strengths. Using the above results combined with the local-density approximation, in this work we propose a method to obtain the strong-coupling expansion for \textcolor{black}{Tan's contact}  of a 1D Bose gas under harmonic confinement to arbitrary order. We also use the conjecture for the ground-state energy to obtain \textcolor{black}{Tan's contact} for all interaction regimes, from weak to strong coupling, as well as the lowest-order term in the weak-coupling expansion.

\section{Local-density approximation for \textcolor{black}{Tan's contact}  of a 1D Bose gas}
\label{sec:1}

We consider $N$ bosonic atoms of mass $m$ at zero temperature, confined by a tight atomic waveguide to a one-dimensional geometry. The atoms interact via a contact potential $v(x-x')=g \delta (x-x')$, where $g$ is the one-dimensional interaction constant, related to the three-dimensional scattering length of the atoms \cite{Olshanii1998}. In the longitudinal direction the atoms are further confined by an external harmonic confinement $ V_{ext}(x)=m \omega_0^2 x^2/2$, describing the optical or magnetic trapping present in the experiments with ultracold atoms. The Hamiltonian of the system reads
\begin{equation}
H=\sum_{j=1}^N \left[ \frac{-\hbar^2}{2m}\frac{\partial^2}{\partial x_j^2} + V_{ext}(x_j)+g\sum_{\ell>j} \delta(x_j-x_\ell)\right].
\end{equation}
In the homogeneous system, the interaction strength is measured in terms of the dimensionless parameter $\gamma=mg/(\hbar^2 \rho)$, where $\rho$ is the average linear density of the homogeneous gas. In the case of harmonic confinement, introducing the harmonic-oscillator length $a_{ho}\!=\!\sqrt{\hbar/(m\omega_0)}$ and the one-dimensional scattering length $a_{\mathrm{1D}}=- 2 \hbar^2/(g m)$, the corresponding dimensionless parameter is $\alpha_0= 2 a_{ho}/(|a_{\mathrm{1D}}|\sqrt{N})$ \cite{Menotti}.

According to Tan's sweep relation, \textcolor{black}{Tan's contact}  in 1D is related to the ground-state energy $E$ of the gas according to \cite{BarthZwerger}
\begin{equation}
C=-\frac{m^2}{\pi \hbar^4} \frac{\partial E}{\partial (1/g)}.
\label{eq:defC}
\end{equation}
In order to determine \textcolor{black}{Tan's contact}  of the harmonically-confined gas, we employ the density-functional approach developed in \cite{Decamp2016}. In detail, we define a functional $E[\rho]$ of the density $\rho(x)$ which, in the local-density approximation, reads
\begin{equation}
E[\rho]=\int dx \, \left[\epsilon(\rho) + (V_{ext}(x)-\mu) \rho(x)\right],
\label{eq:LDA}
\end{equation}
where $\epsilon$ is the ground-state energy density of the homogeneous gas. The ground-state density profile is obtained  by minimizing  the energy functional, i.e. setting $\delta E/\delta \rho\!=\!0$. This yields an implicit equation for the density profile,
\begin{equation}
\frac{3}{2} \frac{\hbar^2}{m} \rho^2 e(\gamma)- \frac{ g \rho}{2}  \frac{\partial e}{\partial \gamma} =\mu -V_{ext}(x)=\mu\, \left(1-\frac{x^2}{R_{TF}^2}\right),
\label{eq:rho}
\end{equation}
where $R_{TF}\!=\!\sqrt{2\mu/(m\omega_0^2)}$ is the Thomas-Fermi radius in a harmonic trap, and the dimensionless average ground-state energy per particle $e$ is such that $\epsilon(\rho)\!=\!\frac{\hbar^2}{2 m} \rho^3 e(\gamma)$. The chemical potential $\mu$ is fixed by imposing the normalization condition $N=\int dx \,\rho(x)$.

Combining Eqs.~(\ref{eq:defC}) and (\ref{eq:LDA}), together with the inhomogeneous density profile (\ref{eq:rho}), we obtain \textcolor{black}{Tan's contact} within the LDA:
\begin{equation}
\label{CLDA}
C_{LDA}=g^2 \frac{m^2}{2 \pi \hbar^4} \int dx \, \rho^2(x) \left.\frac{\partial e}{\partial \gamma}\right|_{\rho=\rho(x)}.
\end{equation}
This expression readily generalizes the known result for the homogeneous system (see e.g. Ref.~\cite{Rigol2015}):
\begin{equation}
\label{Chom}
C=g^2 \frac{m^2}{2 \pi \hbar^4} L \rho^2 \frac{\partial e}{\partial \gamma}.
\end{equation}
In practice, since $e(\gamma)$ is not known analytically for the Lieb-Liniger model, one needs to rely on an approximation scheme to obtain \textcolor{black}{Tan's contact}  in the homogeneous case, and in the trap within the LDA. We proceed by detailing procedures based on various approximations that allow, in principle, to reach excellent accuracy over the whole interaction range.

\section{Scaling relations and methods}

In this Section, we combine the various relations obtained above to systematically compute \textcolor{black}{Tan's contact}  with increasing accuracy. For the energy density $\epsilon(\rho)$, we take the solution stemming from the Lieb-Liniger Bethe Ansatz solution, and use various schemes to estimate $e(\gamma)$, i.e. a strong-coupling expansion
\begin{equation}
\label{estrong}
e(\gamma)=\sum_{k=0}^{+\infty}\frac{a_k}{\gamma^k},
\end{equation}
where $\{a_k\}_{k\geq 0}=\{\pi^2/3, -4\pi^2/3, 4\pi^2, \dots\}$ is currently known up to order $20$ \cite{Lang2016}, as well as the conjecture 
\begin{equation}
\label{econj}
e(\gamma)=\frac{\gamma^2}{3}\sum_{k=0}^{+\infty}\frac{\pi^{2k+2}P_k(\gamma)}{(2+\gamma)^{3k+2}},
\end{equation}
where $\{P_k(X)\}_{k\geq 0}=\{1, \frac{32}{15},-\frac{96}{35}X+\frac{848}{315},\dots\}$ are polynomials explicitly known up to order $6$  \cite{Lang2016}. Expression (\ref{econj}) is very close to the exact numerical Bethe Ansatz solution for a wide range of interaction strengths. It is also a priori possible to rely on a conjectural expansion in the weakly-interacting regime,
\begin{eqnarray}
\label{eweak}
e(\gamma)=\sum_{k=0}^{+\infty}a_k'\gamma^{1+k/2},
\end{eqnarray}
where $\{a_k'\}_{k\geq 0}=\{1,-\frac{4}{3\pi},\frac{1}{6}\!-\!\frac{1}{\pi^2}, \dots\}$ is \textcolor{black}{analytically} known up to second order at the time being \cite{Widom}. \textcolor{black}{Higher-order terms were obtained numerically in a very recent work and the exact value of a few of them has been guessed \cite{Prolhac}.}

\subsection{Strong-coupling expansion for \textcolor{black}{Tan's contact} }
\label{sec:strongc}
We first derive the strong-coupling expansion of \textcolor{black}{Tan's contact}  for a harmonically trapped gas, based on  the strong-coupling expansion Eq.~(\ref{estrong}) for the ground state-energy of the homogeneous system. Combining Eq.~(\ref{eq:rho}), the normalization condition, and the relation $g\!=\!\hbar \omega_0 a_{ho}\sqrt{N}\alpha_0$, the natural rescaled variables are $\overline{\rho}\equiv \rho \, a_{ho}/\sqrt{N}$, $\overline{\mu}\equiv \mu/(N\hbar\omega_0)$ and $\overline{x}\equiv x/R_{TF}$, whereupon we obtain the following set of equations:
\begin{eqnarray}
\label{eq1strong}
\frac{1}{2}\sum_{k=0}^{+\infty}\frac{A_k}{\alpha_0^k}\overline{\rho}^{k+2}(\overline{x},\alpha_0)=(1-\overline{x}^2)\overline{\mu}(\alpha_0),
\end{eqnarray}
where $A_k\equiv (k+3)a_k$, and also
\begin{eqnarray}
\label{eq2strong}
1\!=\!\sqrt{2\overline{\mu}}\int_{-1}^1d\overline{x}\,\overline{\rho}(\overline{x}).
\end{eqnarray}
A straightforward approach to solve the above equations proceeds as follows. One truncates the series in Eq.~(\ref{eq1strong}) to order $n$. Using the values for the  coefficients $\{a_k\}_{k\leq n}$ from Eq.~(\ref{estrong}), one can express $\overline{\rho}$ as the root of a $(n\!+\!2)^{th}$-degree polynomial as a function of $\overline{\mu}$ and inject into Eq.~(\ref{eq2strong}) to find $\overline{\mu}$, and thus $\overline{\rho}$ explicitly, consistently to order $n$ by expanding the obtained solution in $1/\alpha_0$. This approach, when carried analytically, suffers from two major drawbacks. First, it is limited to $n\!=\!2$ since in general, polynomials of order strictly higher than $4$ can not be solved by radicals. Second, among the $n$ possible solutions, one has to select the physical one. Using Cardan's method for $n\!=\!1$ and Ferrari's method for $n\!=\!2$, putting the roots of the polynomials in a trigonometric form for simplicity, we have used this approach to compute the strong-coupling expansion of \textcolor{black}{Tan's contact}  to second order in $1/\alpha_0$. This has also been useful to benchmark the more general method described just below. 

In order to obtain higher-order terms in the strong-coupling expansion, we have 
developed a more efficient procedure, which allows to systematically compute the expansion of \textcolor{black}{Tan's contact}  to arbitrary order. The latter relies on the following expansions: $\overline{\mu}=\sum_{k=0}^{+\infty}\frac{c_k}{\alpha_0^k}$, and $\overline{\rho}(\overline{x})=\sum_{j=0}^{+\infty}\frac{b_j}{\alpha_0^j}f_j(\overline{x})$, where $\{f_j\}_{j\geq 0}$ is a set of unknown functions, injected into Eqs.~(\ref{eq1strong}, \ref{eq2strong}). By analysis, we find that the consistency condition at a given order converts into $b_jf_j(\overline{x})=\sum_{m=0}^jb_{mj}(1-\overline{x}^2)^{(m+1)/2}$, where $\{b_{mj}\}$ are unkwown coefficients of an upper triangular matrix. Then, synthesis yields:
\begin{eqnarray}
\label{eq1strongfin}
\frac{1}{2}\sum_{k=0}^{+\infty}\frac{A_k}{\alpha_0^k}\left(\sum_{j=0}^{+\infty}\frac{1}{\alpha_0^j}\sum_{m=0}^jb_{mj}(1-\overline{x}^2)^{(m+1)/2}\right)^{k+2}=\left(1-\overline{x}^2\right)\sum_{k=0}^{+\infty}\frac{c_k}{\alpha_0^k}
\end{eqnarray}
and
\begin{eqnarray}
\label{eq2strongfin}
1=32\sum_{k=0}^{+\infty}\frac{c_k}{\alpha_0^k}\left[\sum_{j=0}^{+\infty}\frac{1}{\alpha_0^j}\sum_{m=0}^jb_{mj}2^mB\left(\frac{m+3}{2},\frac{m+3}{2}\right)\right]^2,
\end{eqnarray}
where $B$ is the Euler Beta function. Equations (\ref{eq1strongfin}) and (\ref{eq2strongfin}) are the final set of equations. As can be seen, solving the system truncated to order $n$ requires the solution at all lower orders, thus the procedure becomes increasingly lengthy. Moreover, at each step Eq.~(\ref{eq1strongfin}) splits into $n\!+\!1$ independent equations, obtained by equating the coefficients of $(1-\overline{x}^2)^{(1+m)/2}_{m=0,\dots,n}$ in the LHS and RHS. One thus needs to solve a system of $n\!+\!2$ equations to obtain $c_n$ and $\{b_{mn}\}_{m=0,\dots, n}$, but fortunately, $n$ of them, giving $b_{mj}$, $m\geq 1$, are fully decoupled.

As a final step, using Eq.~(\ref{CLDA}), we obtain \textcolor{black}{Tan's contact} . In natural units imposed by the scaling, i.e. taking $\overline{C}_{LDA}=C_{LDA} a_{ho}^3/N^{5/2}$, we have
\begin{eqnarray}
\overline{C}_{LDA}\!=-\frac{1}{\pi\sqrt{2}}\sqrt{\sum_{k'=0}^{+\infty}\frac{c_{k'}}{\alpha_0^{k'}}}\sum_{k=0}^{+\infty}\frac{B_k}{\alpha_0^k}\int_{-1}^1\!d\overline{x}\left(\sum_{j=0}^{+\infty}\frac{1}{\alpha_0^j}\sum_{m=0}^jb_{mj}(1-\overline{x}^2)^{(m+1)/2}\!\right)^{k+4}\!\!\!\!,
\end{eqnarray}
where $B_k\!=\!(k\!+\!1)a_{k+1}$. Note that the value is positive because $a_1\!<\!0$. At order $n$, the condition $k'+k+j'=n$, where $j'$ is the power of $\alpha_0$ in the integrand, shows that the coefficient of order $n$ is a sum of $\binom{n+2}{n}$ integrals. One of them involves $a_{n+1}$, so one needs to know the function $e(\gamma)$ to order $n\!+\!1$ in $1/\gamma$ to obtain the expansion of the Tan's contact to order $n$ in $1/\alpha_0$.

\subsection{Tan's contact at arbitrary interactions from the conjecture}
To check the validity of the strong-coupling method, we have derived another resolution scheme, from Eq.(\ref{econj}). After rescaling and straightforward algebra, one obtains the mixed form
\begin{eqnarray}
\frac{\alpha_0^2}{6}\sum_{k=0}^{+\infty}\frac{\pi^{2k+2}}{(\gamma+2)^{3k+3}}\left[(\gamma+2)P_k(\gamma)-\gamma(\gamma+2)P_k'(\gamma)+3k+2\right]=\overline{\mu}(\alpha_0)(1-\overline{x}^2).
\label{muloc}
\end{eqnarray}
Once truncated at order $n$, it yields a polynomial in $1/(\gamma+2)$ 
whose roots are found numerically. 
We proceed as follows. We start from a guessing value of $\overline{\mu}$
and from Eq. (\ref{muloc}) we obtain $\overline{\rho}(\overline{\mu},\overline{x})$.
We integrate the density and we use the difference $N-\int\rho\, dx$ 
as control parameter of the accuracy of the initial value
for $\overline{\mu}$. A findroot subroutine exploits this contol parameter
to converge to the $\overline{\mu}$ value that ensures
the density normalization to $N$ bosons.
Once the correct density is computed, we get $C_{LDA}$ from Eq. (\ref{CLDA}).

\subsection{Weak-coupling expansion for \textcolor{black}{Tan's contact} }
\label{sec:weakc}
At weak iteractions, we derive an expression for \textcolor{black}{Tan's contact}  using the weak-coupling expansion Eq.~(\ref{eweak}). Using the same notations as above, we obtain
\begin{eqnarray}
\sum_{k=0}^{+\infty}\frac{a_k'}{4}(4-k)\overline{\rho}^{\frac{2-k}{2}}(\overline x,\alpha_0)\alpha_0^{\frac{k+2}{2}}=\left(1-\overline{x}^2\right)\overline{\mu}(\alpha_0).
\end{eqnarray}
On the other hand, the normalization condition is given by Eq.~(\ref{eq2strong}) as before. Here, it is not obvious to what order one should truncate the expressions to obtain a consistent expansion at given order, nor to find the variable in which to expand, as can be seen by evaluating the first orders. Considering only the $k\!=\!0$ term in the sum, one easily finds $\overline{\rho}(\overline{x})=\left(9/32\right)^{1/3}(1-\overline{x}^2)/\alpha_0^{1/3}$ and $\overline{\mu}(\alpha_0)=\left(9/32\right)^{1/3}\alpha_0^{2/3}$.  \textcolor{black}{The expansion to next order is problematic. If one retains terms up to $k\!=\!1$, corresponding to the Bogoliubov approximation,  since the coefficient $a'_1$ is negative, the equation of state at sufficiently large density becomes negative \cite{LiebLiniger}. Then it is not possible to use it to perform the local-density approximation.} One may also recall that the local-density approximation breaks down at very weak interactions, where it is not accurate to neglect the quantum tails in the density profile. In this regime, a different scaling parameter becomes relevant \cite{Zvonarev2014,Olshanii2014}.

\section{Results for \textcolor{black}{Tan's contact} }
Following the approach presented in Sec.~\ref{sec:strongc}, the strong-coupling expansion reads:
\begin{eqnarray}
\label{eq:strongc}
&&\overline{C}_{LDA}=\frac{128\sqrt{2}}{45 \pi^3}+\frac{1}{\alpha_0}\left(-\frac{8192}{81\pi^5}+\frac{70}{9\pi^3}\right)+\frac{\sqrt{2}}{\alpha_0^2}\left(\frac{131072}{81\pi^7}-\frac{30656 }{189\pi^5}-\frac{4096}{525 \pi^3}\right)\nonumber\\
&&+\frac{1}{\alpha_0^3}\left(-\frac{335544320}{6561  \pi^9}+\frac{4407296}{729\pi^7}+\frac{872701}{2025\pi^5}-\frac{112}{3\pi^3}\right)\nonumber\\
&&+\frac{\sqrt{2}}{\alpha_0^4}\left(\frac{47982837760 }{59049  \pi^{11}}-\frac{717291520 }{6561 \pi^9}-\frac{108494512 }{10935 \pi^7}+\frac{2112512}{1701 \pi^5}+\frac{65536}{2205\pi^3}\right).
\end{eqnarray}
This expression agrees with \textcolor{black}{the zero order one obtained in \cite{Olshanii} and with} the one derived for a $\kappa$-component balanced spinful Fermi gas in \cite{Decamp2016} to order $2$: in fact one may find the bosonic result by taking the limit of very large number of fermionic components $\kappa\to\infty$ \cite{Yang2011}.

The weak-coupling expansion obtained from Sec.~\ref{sec:weakc} reads:
\begin{equation}
\label{eq:weakc}
\overline{C}_{LDA}=\frac{1}{5\pi}\left(\frac{3}{2}\right)^{2/3}\alpha_0^{5/3},
\end{equation}
\textcolor{black}{in agreement with \cite{Olshanii}}.

\begin{figure}
\resizebox{0.85\columnwidth}{!}{%
\includegraphics{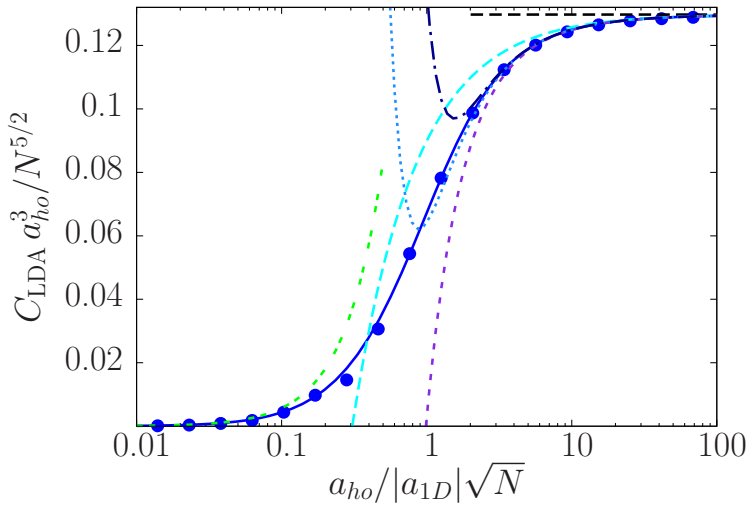}}
\caption{Scaled Tan's contact for a 1D Bose gas (in units of $N^{5/2}/a_{ho}^3$) as a function of the dimensionless interaction strength $\alpha_0/2=a_{ho}/(|a_{\mathrm 1D}|\sqrt{N})$. Results from the strong-coupling expansion  (\ref{eq:strongc}): Tonks-Girardeau (horizontal long-dashed line, black),  $1^{st}$ order correction (long dashed, cyan), $2^{nd}$ order correction (short-dashed, purple), $3^{rd}$ order correction (dotted, light blue), $4^{th}$ order correction (dot-dashed, dark blue). Results at arbitrary interactions:   conjecture (\ref{econj}) to order six  (blue dots),  exact equation of state (data from Ref.\cite{Olshanii}, continuous, blue). We also show  the weak-coupling expansion (\ref{eq:weakc}) (double dashed, green).}
\label{fig:1}       
\end{figure}

Figure \ref{fig:1} summarizes our results for \textcolor{black}{Tan's contact}. Notice that, although the contact is scaled by the overall factor $N^{5/2}/a_{ho}^3$\textcolor{black}{, it still} depends on the number of particles through the factor $\alpha_0/2=a_{ho}/|a_{1D}|\sqrt{N}$.
First, we notice that the results based on the conjecture are extremely close to the ones obtained from the full solution of the Bethe-Ansatz equation of state in Ref.\cite{Olshanii}. Second, by comparing the strong-coupling expansion with the results of the full calculation, we notice that the expansion (\ref{eq:strongc}) is valid down to $\alpha_0/2\simeq 3$, and provides an useful analytical expression for Tan's contact in harmonic trap. In order to accurately describe the regime of lower interactions one would need a considerable number of terms in the strong-coupling expansion. The same is  true for the series expansion for the ground-state energy of the homogenenous gas \cite{Lang2016}. The use of the conjecture (\ref{econj}) is then a valuable alternative with respect to solving numerically the Bethe-Ansatz integral equations, the weak coupling expansion being applicable only for very weak interactions $\alpha_0/2\simeq 0.05$.

\section{Conclusions and outlook}
In conclusion, in this work we have determined  \textcolor{black}{Tan's contact} for a harmonically-trapped 1D Bose gas. In particular, using an asymptotic expansion of the ground-state energy of the homogeneous system and the local-density approximation, we have developed a general method to obtain a strong-coupling expansion of \textcolor{black}{Tan's contact} to arbitrary order, and have provided the coefficients for the expansion up to order four. We have tested it against a calculation of \textcolor{black}{Tan's contact} based on a conjecture of the ground-state energy at arbitrary interactions \cite{Lang2016} as well as to the full solution of the Bethe-Ansatz equations provided in Ref.\cite{Olshanii}. The strong-coupling expansion yields an accurate expression at large interactions, but requires a considerable  number of terms to obtain good accuracy at intermediate interactions. In this parameter regime it is then useful to apply  the method based on the conjecture for the equation of state. In outlook, the method presented in this work could be used to calculate higher-order terms of the strong coupling expansion \textcolor{black}{with the aim of resumming the terms as was done to obtain the conjectural expression for the homogeneous gas}. The local-density approximation could be tested by comparing with ab-initio numerical simulations. It would be interesting to generalize this method to the case of multicomponent 1D gases as well as to finite temperature, beyond the infinitely repulsive Tonks-Girardeau limit of Refs.~\cite{Vignolo2013,Rigol2015}.

\section{Acknowledgements}
We thank M. Albert, M. Rizzi and J. Decamp for discussions. We thank V. Dunjko and M. Olshanii for providing us the data of Ref.\cite{Olshanii}. We acknowledge financial support from the ANR projects Mathostaq (ANR-13-JS01-0005-01) and SuperRing (ANR-15-CE30-0012-02). AM wishes to dedicate this work to Roger  Maynard, which follows his continuous encouragement and enthusiasm.

\end{document}